\documentclass[prb,twocolumn,showpacs,floatfix]{revtex4-1}
\usepackage{amsmath}
\usepackage{amssymb}
\usepackage{graphicx}
\usepackage[english]{babel}
\usepackage{latexsym}
\usepackage{graphics}
\usepackage{subfigure}

\begin{document}

\title{Robustness of Majorana Modes and Minigaps in a Spin-Orbit-Coupled
Semiconductor-Superconductor Heterostructure}
\author{Li Mao}
\author{Chuanwei Zhang}
\thanks{Email: cwzhang@wsu.edu}
\affiliation{Department of Physics and Astronomy, Washington State University, Pullman,
WA, 99164 USA}
\begin{abstract}
We study the robustness of Majorana zero energy modes and minigaps of
quasiparticle excitations in a vortex by numerically solving
Bogoliubov-deGennes equations in a heterostructure composed of an \textit{s}%
-wave superconductor, a spin-orbit-coupled semiconductor thin film, and a
magnetic insulator. This heterostructure was proposed recently as a platform
for observing non-Abelian statistics and performing topological quantum
computation. The dependence of the Majorana zero energy states and the
minigaps on various physics parameters (Zeeman field, chemical potential,
spin-orbit coupling strength) is characterized. We find the minigaps depend
strongly on the spin-orbit coupling strength. In certain parameter region,
the minigaps are linearly proportional to the \textit{s}-wave
superconducting pairing gap $\Delta _{s}$, which is very different from the $%
\Delta _{s}^{2}$ dependence in a regular \textit{s- }or\textit{\ p}-wave
superconductor. We characterize the zero energy chiral edge state at the
boundary and calculate the STM signal in the vortex core that shows a
pronounced zero energy peak. We show that the Majorana zero energy states
are robust in the presence of various types of impurities. We find the
existence of impurity potential may increase the minigaps and thus benefit
topological quantum computation.
\end{abstract}

\maketitle

\section{Introduction}

Topological quantum computation (TQC) \cite{NayakRMP,SarmaPT}, where quantum
information is processed using a decoherence-free subspace guaranteed by
topological order, is a revolutionary new alternative to conventional
quantum computing. In TQC, quantum information is encoded in certain
nonlocal, topological, degrees of freedom of the underlying physical system (%
\textit{i.e.}, hardware) that do not couple to weak local noise. This
special hardware, called 'non-Abelian topological matter', has been proposed
to exist in certain classes of two-dimensional (2D) strongly-correlated
systems, such as the $\nu =5/2$ fractional quantum Hall (FQH) system \cite%
{Moore91,Nayak96,Read00,Sarma05,Stern06,Bonderson06,Rosenow08}, chiral
\textit{p}-wave superconductor/superfluid \cite%
{Ivanov01,Stern04,Sarma06,Tewari07,Tewari072,Tewari08,Cheng05,Gurarie05,Gurarie07,Melo05,Zhang072}%
, and some artificial states of cold atoms in optical lattices \cite%
{Kitaev06,Duan03,Zhang07}. In these systems, the ground state wave function
is a linear combination of states from a degenerate subspace, and a
pair-wise exchange of the particle coordinates unitarily rotates the ground
state wave function in this subspace. Therefore, the exchange statistics of
the particles is given by a multi-dimensional unitary matrix representation
(as opposed to just a phase factor for bosons and fermions) of the 2D braid
group, and the statistics is non-Abelian.

Despite the tremendous technological potential, non-Abelian topological
matter is rare in nature and generally hard to observe in experiments \cite%
{NayakRMP}. To circumvent this problem, there has been considerable
interests recently for exploring the possibility of `designing' non-Abelian
topological order in the fertile laboratory of the cold atom systems \cite%
{Zhang08,Sato09} and the regular solid state materials \cite%
{Fu08,Fu09,Akhmerov09,Fu10,Sau10,Stanescu10,Lee09,Qi10,Tewari10,Alicea10,Sau102,Sau103,Sau104,Sau105,Linder}%
. Two important resources for the emergence of non-Abelian statistics in a
correlated matter are (a) chirality of the constituent particles and (b)
superconducting order. In a composite system, these two basic ingredients of
topological order may arise from two different physical effects (for
instance, spin-orbit coupling and \textit{s}-wave superconductivity,
respectively) to design a non-Abelian quantum state. This strategy allows us
to use \textit{s}-wave superconductors/superfluids, which are much more
abundant in nature and much less sensitive to disorder effects than their
\textit{p}-wave counterparts.

One important recent progress along this direction was the proposed solid
state heterostructure composed of an electron-doped semiconductor thin film,
an \textit{s}-wave superconductor, and a magnetic insulator as a non-Abelian
platform for TQC \cite{Sau10}. In this heterostructure, the superconducting
pairing order is induced on the semiconductor thin film from the \textit{s}%
-wave superconductor through the superconducting proximity effect, and the
chirality of particles is supplied by the Rashba spin-orbit coupling in the
semiconductor \cite{Rashba}. With a set of vortices in the heterostructure,
there exist one Majorana zero energy state of quasiparticle excitations in
each vortex core. These zero energy states are the topological, degenerate,
ground states following non-Abelian statistics and can be used to perform
TQC.

In addition to the existence of Majorana zero energy modes, another key
ingredient for the physical implementation of TQC is that the degenerate
ground state subspace must be separated from other non-topological excited
states by an energy gap, so that finite temperature cannot populate the
excited states and ruin the topological properties of the system. Recently,
it was shown\cite{Akhmerov10}, by constructing a new type of Majorana
operators that contain both zero energy and excited states, that the
topological braiding statistics of Majorana fermions may still preserve even
in the presence of non-topological excitations. However, the minigap is
still important because the signal strengths of measurements will be reduced
significantly when the temperature is comparable with the minigap where the
number of non-topological excitations become significant. The magnitude of
the energy gap directly determines the operating temperature of the
underlying physical system as a realistic TQC platform and is of critical
importance. Furthermore, both Majorana zero energy states and the magnitude
of the gap must be robust in the presence of various impurities. Although
the existence of Majorana zero energy states in the vortex core in a
heterostructure has been proved by analyzing the zero energy solution of the
Bogoliubov-deGennes (BdG) equation, the magnitudes of the minimum energy
gaps and their robustness against impurities have not been addressed.

In this paper, we numerically solve the BdG equation for a vortex in the
heterostructure and calculate the minimum energy gap (minigap) between the
zero energy state and the first quasiparticle excited state in the vortex
core. In the simulation, the proximity-induced $s$-wave superconducting
pairing gap $\Delta _{s}$ in the semiconductor is obtained from the
self-consistent solution of the BdG equations for a pure $s$-wave
superconductor. The main results are summarized as follows:

1) The full numerical simulation of the BdG equations confirms that the
Majorana zero energy states exist only in the parameter region $V_{z}>\sqrt{%
\mu ^{2}+\Delta _{s}^{2}}$, which agrees with previous results obtained from
an approximate analytical approach. Here $V_{z}$ is the perpendicular Zeeman
field induced by the proximity contact with the magnetic insulator, $\mu $
is the chemical potential of the electron gas.

2) The dependence of the minigap $E_{g}$ on various parameters $\left(
V_{z},\mu ,\alpha \right) $ is characterized. Here $\alpha $ is the strength
of the Rashba spin-orbit coupling in the semiconductor. We find $E_{g}$
depends strongly on $\alpha $. In certain parameter region, $E_{g}$ is $\sim
\Delta _{s}$, instead of $\sim \Delta _{s}^{2}$ for a regular chiral $p$%
-wave superconductor/superfluid \cite{gap}.

3) An analytical theory is developed to explain the properties of the zero
energy chiral edge states at the boundary.

4) The scanning tunneling microscopy (STM) signals around the vortex show
pronounced peaks at the zero energy and the minigap, thus can be used to
detect the zero energy modes and the minigaps in experiments.

5) We show that Majorana zero energy modes are robust in the presence of
various impurity potentials. Surprisingly, we find the existence of impurity
potential in the vortex core may increase the magnitudes of the minigaps,
and thus may be useful for the physical implementation of TQC.

The paper is organized as follows: Section II lays out the BdG equation for
a vortex in a semiconductor-superconductor heterostructure. Section III
discusses the parameter dependence of the zero energy states and the minigap
$E_{g}$. In Sec. IV, we discuss the robustness of the zero energy states and
the minigaps in the presence of realistic impurities. Section V consists of
conclusions. The details about the numerical approach to the BdG equations
are presented in Appendix A. The finite size effect in the BdG equations is
discussed in Appendix B. In Appendix C, we discuss the zero energy chiral
edge modes at the boundary.

\section{BdG equations for a vortex}

\begin{figure}[t]
\includegraphics[width=1.0\linewidth]{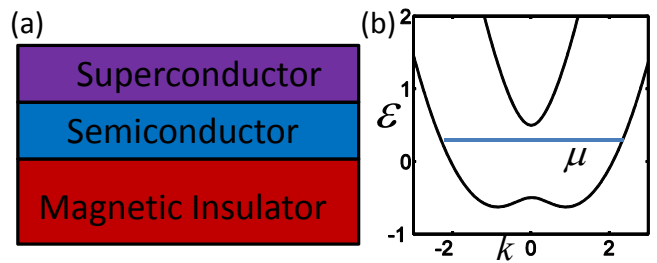} \vspace{0pt}
\caption{(Color online) (a) An illustration of the structure of the
spin-orbit coupled semiconductor-superconductor heterostructure. (b) The
single particle energy spectrum in a spin-orbit coupled semiconductor.}
\label{system}
\end{figure}

The physical system we consider is a heterostructure composed of an \textit{s%
}-wave superconductor, an electron-doped semiconductor thin film, and a
magnetic insulator (Fig. \ref{system}a). The dynamics of electrons in the
semiconductor are described by a single particle effective Hamiltonian
\begin{equation}
H_{0}=\frac{p^{2}}{2m^{\ast }}-\alpha \left( p_{x}\sigma _{y}-p_{y}\sigma
_{x}\right) +V_{z}\sigma _{z}-\mu ,  \label{singleHam}
\end{equation}%
where $m^{\ast }$ is the conduction-band effective mass of electrons, $\mu $
is the chemical potential, $\alpha $ is the strength of the Rashba
spin-orbit coupling, $V_{z}$ is a perpendicular Zeeman field induced by the
proximity contact with the magnetic insulator. $\sigma _{i}$ are the Pauli
matrices for the electron spins.

The Hamiltonian yields two spin-orbit bands (Fig. \ref{system}b) with energy
dispersions%
\begin{equation}
\varepsilon _{\pm }=\frac{k^{2}}{2m^{\ast }}-\mu \pm \sqrt{\alpha
^{2}k^{2}+V_{z}^{2}}  \label{singleene}
\end{equation}%
in a uniform system. Henceforth we set $\hbar =1$. A finite energy gap $%
2V_{z}$ is opened at $k=0$ for a nonzero $V_{z}$. When the chemical
potential lays in the gap, electrons only occupy the lower spin-orbit band
at a low temperature.

The mean field Hamiltonian of an $s$-wave superconductor can be written as
\begin{equation}
\hat{H}=\int d\mathbf{r}\left(
\begin{array}{cc}
a_{\uparrow }^{\dag }\left( \mathbf{r}\right) & a_{\downarrow }\left(
\mathbf{r}\right)%
\end{array}%
\right) \left(
\begin{array}{cc}
\hat{H}_{s} & \Delta (\mathbf{r}) \\
\Delta ^{\ast }(\mathbf{r}) & -\hat{H}_{s}%
\end{array}%
\right) {\binom{a_{\uparrow }\left( \mathbf{r}\right) }{a_{\downarrow
}^{\dag }\left( \mathbf{r}\right) }}  \label{BdG1}
\end{equation}%
in the Nambu space, where $\hat{H}_{s}=-\nabla ^{2}/2m-E_{F}+U(\mathbf{r})$
is the single particle Hamiltonian, $E_{F}$ is Fermi energy, $U(\mathbf{r})$
is an external potential, $a_{\sigma }\left( \mathbf{r}\right) $ are
annihilation operators of electrons for position eigenfunctions rather than
for momentum eigenfunctions, $\Delta (\mathbf{r})$ is the \textit{s}-wave
pairing order parameter. The Hamiltonian (\ref{BdG1}) can be diagonalized by
the Bogoliubov transformation
\begin{equation}
{\binom{a_{\uparrow }(\mathbf{r})}{a_{\downarrow }^{\dag }(\mathbf{r})}}%
=\sum_{n}\left(
\begin{array}{cc}
u_{n}(\mathbf{r}) & -v_{n}^{\ast }(\mathbf{r}) \\
v_{n}(\mathbf{r}) & u_{n}^{\ast }(\mathbf{r})%
\end{array}%
\right) {\binom{\gamma _{n\uparrow }}{\gamma _{n\downarrow }^{\dag }},}
\label{BdG2}
\end{equation}%
where the wavefunctions $u_{n}(\mathbf{r})$, $v_{n}(\mathbf{r})$ satisfy the
BdG equation
\begin{equation}
\left(
\begin{array}{cc}
\hat{H}_{s} & \Delta (\mathbf{r}) \\
\Delta ^{\ast }(\mathbf{r}) & -\hat{H}_{s}%
\end{array}%
\right) {\binom{u_{n}(\mathbf{r})}{v_{n}(\mathbf{r})}}=E_{n}{\binom{u_{n}(%
\mathbf{r})}{v_{n}(r)}}  \label{BdG3}
\end{equation}%
and the normalization condition%
\begin{equation}
\int d\mathbf{r}[u_{m}^{\ast }(\mathbf{r})u_{n}(\mathbf{r})+v_{m}^{\ast }(%
\mathbf{r})v_{n}(\mathbf{r})]=\delta _{mn}.  \label{norm}
\end{equation}%
The order parameter $\Delta (\mathbf{r})$ can be determined self-consistently%
\cite{Gygi} through the relation $\Delta (\mathbf{r})=g\sum_{n}u_{n}(\mathbf{%
r})v_{n}^{\ast }(\mathbf{r})$. Here $g$ is the effective electron-electron
interaction strength in the superconductor.

The proximity effect between the \textit{s}-wave superconductor and the
semiconductor induces an effective superconducting pairing for electrons in
the semiconductor described by the Hamiltonian%
\begin{equation}
H_{p}=\int d\mathbf{r}\left\{ \Delta _{s}(\mathbf{r})c_{\uparrow }^{\dag
}\left( \mathbf{r}\right) c_{\downarrow }^{\dag }\left( \mathbf{r}\right) +%
\text{H.c}.\right\} ,  \label{pairing}
\end{equation}%
where $c_{\sigma }^{\dag }\left( \mathbf{r}\right) $ are the creation
operators for electrons, and $\Delta _{s}(\mathbf{r})$ is the
proximity-induced effective \textit{s}-wave pairing gap in the semiconductor
thin film. Because of the Rashba spin-orbit coupling in the semiconductor,
the single particle Hamiltonian $\hat{H}_{s}$ in the BdG equation (\ref{BdG3}%
) should now be replaced with $H_{0}$ in (\ref{singleHam}). The BdG equation
written in the Nambu spinor basis becomes%
\begin{equation}
\left(
\begin{array}{cc}
H_{0} & \Delta _{s}(\mathbf{r}) \\
\Delta _{s}^{\ast }(\mathbf{r}) & -\sigma _{y}H_{0}^{\ast }\sigma _{y}%
\end{array}%
\right) \Phi _{n}(\mathbf{r})=E_{n}\Phi _{n}(\mathbf{r}),  \label{BdG4}
\end{equation}%
where $\Phi _{n}(\mathbf{r})=\left[ u_{n\uparrow }(\mathbf{r}%
),u_{n\downarrow }(\mathbf{r}),v_{n\downarrow }(\mathbf{r}),-v_{n\uparrow }(%
\mathbf{r})\right] ^{T}$ is the quasiparticle wavefunction. The Bogoliubov
quasiparticle operator is
\begin{equation}
\gamma _{n}^{\dag }=\int d\mathbf{r}\sum_{\sigma }\left[ u_{n\sigma }(%
\mathbf{r})c_{\sigma }^{\dag }\left( \mathbf{r}\right) +v_{n\sigma }(\mathbf{%
r})c_{\sigma }\left( \mathbf{r}\right) \right] .  \label{Bog}
\end{equation}

\begin{figure}[t]
\includegraphics[width=0.9\linewidth]{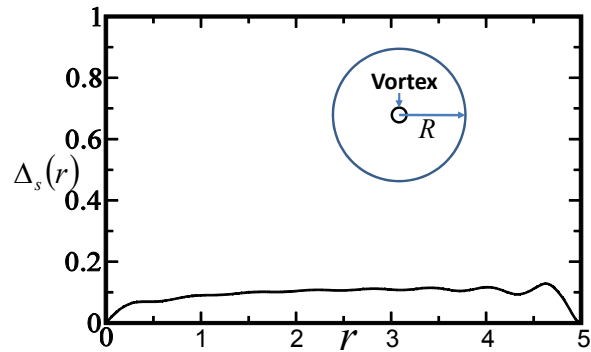} \vspace{-0pt}
\caption{Plot of the \textit{s}-wave pairing gap $\Delta _{s}\left( r\right)
$ from the self-consistent solution of the BdG equation (\protect\ref{BdG3})
for a pure \textit{s}-wave superconductor. }
\label{delta}
\end{figure}

In the presence of a vortex in the semiconductor-superconductor
heterostructure, the order parameter takes the form $\Delta _{s}(r,\theta
)=\Delta _{s}(r)e^{i\theta }$, and the solutions $\left( E_{n},\Phi _{n}(%
\mathbf{r})\right) $ of the BdG equation (\ref{BdG4}) correspond to the
quasiparticle excitation energies and states in the vortex core. For
simplicity of the calculation, we consider a two-dimensional cylinder
geometry with a hard wall at the radius $r=R$ and a single vortex at $r=0$.
This system preserves the rotation symmetry and the BdG equation can be
decoupled into different angular momentum channels indexed by $l$ with the
corresponding spinor wavefunction%
\begin{equation}
\Phi _{n}^{l}(\mathbf{r})=e^{il\theta }\left[ u_{n\uparrow
}^{l}(r),u_{n\downarrow }^{l}(r)e^{i\theta },v_{n\downarrow
}^{l}(r)e^{-i\theta },-v_{n\uparrow }^{l}(r)\right] ^{T}.  \label{wave}
\end{equation}%
Note that the BdG equation has the particle-hole symmetry, therefore if $%
\Phi _{n}^{l}(\mathbf{r})$ is a solution with an energy $E$, then there is
another solution with the energy $-E$ in the angular momentum $-l$ channel.
Henceforth we only consider $E\geq 0$ solutions.

\section{Majorana modes and minigaps}

We numerically solve the BdG equation (\ref{BdG4}) with a vortex and
calculate the quasiparticle excitation energy $E_{n}$ and wavefunction $\Phi
_{n}(\mathbf{r})$ for various parameters $\left( V_{z},\mu ,\alpha \right) $%
. In the numerical treatment, the radial wave functions $u_{n\sigma
}^{l}(r),v_{n\sigma }^{l}(r)$ in Eq. (\ref{wave}) are expanded on an
orthogonal basis $\phi _{jl}(r)=\sqrt{2}J_{l}(\beta
_{jl}r/R)/[RJ_{l+1}(\beta _{jl})]$, where $J_{l}(x)$ is the $l$-th order
Bessel function, $\beta _{jl}$ is the $j$-th zero of $J_{l}(x)$. This basis
satisfies the boundary condition $\phi _{jl}(R)=0$ automatically. The BdG
Hamiltonian (\ref{BdG4}) can be written as a matrix form on this basis and
then diagonalized to obtain $E_{n}$ and $\Phi _{n}(\mathbf{r})$ (more
details about the numerical method can be found in Appendix A). Henceforth
we choose $k_{c}^{-1}=k_{F}^{-1}\left( m/m^{\ast }\right) ^{1/2}=5k_{F}^{-1}$
as the length unit and $\eta =\hbar ^{2}k_{c}^{2}/2m^{\ast }=E_{F}$ as the
energy unit, where $k_{F}$ is the Fermi wavevector in the \textit{s}-wave
superconductor, and we adopt an effective mass $m^{\ast }=0.04m$ for
electrons in the conduction band of a semiconductor.
\begin{figure}[b]
\includegraphics[width=0.9\linewidth]{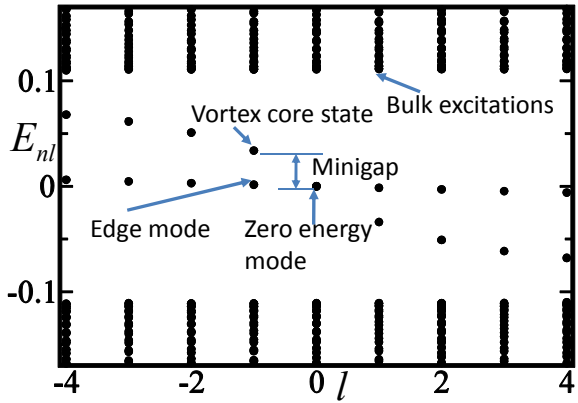} \vspace{-0pt}
\caption{(Color online) Plot of the quasiparticle energies $E_{nl}$ in a
vortex for the angular momentum $l$ from $-4$ to $4$. $\protect\alpha =1$, $%
\protect\mu =0$, and $V_{z}=0.3$.}
\label{band}
\end{figure}

The \textit{s}-wave pairing gap $\Delta _{s}(\mathbf{r})$ in the BdG
equation (\ref{BdG4}) for the semiconductor-superconductor heterostructure
is obtained by solving the BdG equation (\ref{BdG3}) self-consistently for a
pure \textit{s}-wave superconductor\cite{Gygi}, and the resulting $\Delta
_{s}(r)$ is plotted in Fig. \ref{delta}. In the self-consistent procedure,
we first guess a form of $\Delta _{s}(r)$ and inset it into the BdG Eq. (\ref%
{BdG3}), from which we obtain the wavefunction $u_{n}(r)$ and $v_{n}(r)$. $%
\Delta _{s}(r)$ is then self-consistently determined from $u_{n}(r)$ and $%
v_{n}(r)$ through the relation $\Delta _{s}(r)=g\sum_{n}u_{n}(r)v_{n}^{\ast
}(r)$. The new $\Delta _{s}(r)$ is inset into the BdG Eq. (\ref{BdG3}) to
start another cycle of the calculation. The procedure continues until $\Delta
_{s}(r)$ converges. $\Delta _{s}(r)$ is zero at the vortex core, approaches
a uniform value $\Delta _{0}\approx 0.11$ in the bulk, and drops to zero at
the boundary, as expected.
\begin{figure}[t]
\includegraphics[width=1\linewidth]{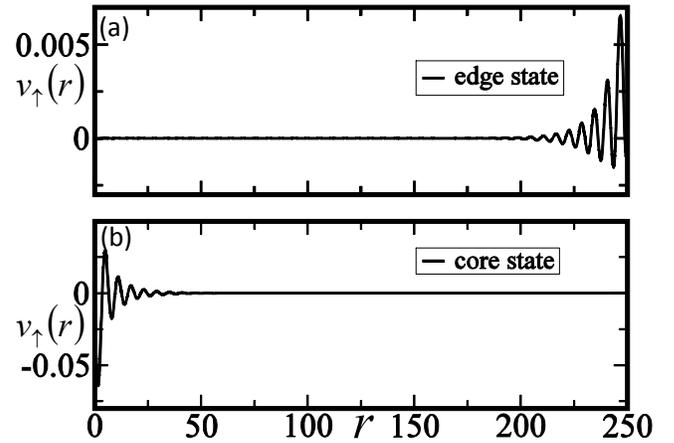} \vspace{-0pt}
\caption{Plot of the wavefunctions $v_{\uparrow }\left( r\right) $ of the
vortex core state and the edge state in the $l=-1$ angular momentum channel.
$\protect\mu =0$, $V_{z}=0.3$, $\protect\alpha =1$.}
\label{edgecorefun}
\end{figure}

The finite size effect can become important in the calculation of the
Majorana zero energy modes and the minigaps in certain parameter region in
the numerical simulation of the BdG equation. More details about the finite
size effect can be found in Appendix B. Here we choose a large radius $R=250$
of the cylinder to suppress the finite size effect. In practice, solving the
BdG equation self-consistently to obtain $\Delta _{s}(\mathbf{r})$ for a
pure \textit{s}-wave superconductor with a large radius\ $R=250$ is very
time costing. Since $\Delta _{s}(\mathbf{r})$ approaches the uniform bulk
value $\Delta _{0}$ within a finite distance from the vortex core, we choose
a pairing gap $\Delta _{s}(r)$ based on the numerical result $\bar{\Delta}%
_{s}(r)$ from the $\bar{R}=5k_{c}^{-1}$ calculation (Fig. \ref{delta}),
\textit{i.e.}, $\Delta _{s}(r)=\bar{\Delta}_{s}(r)$ for $0\leq r\leq 2.8$, $%
\Delta _{s}(r)=\Delta _{0}$ for $2.8<r\leq R-2.2$, and $\Delta _{s}(r)=\bar{%
\Delta}_{s}(r-R+5)$ for $R-2.2<r\leq R$.
\begin{figure}[t]
\includegraphics[width=1\linewidth]{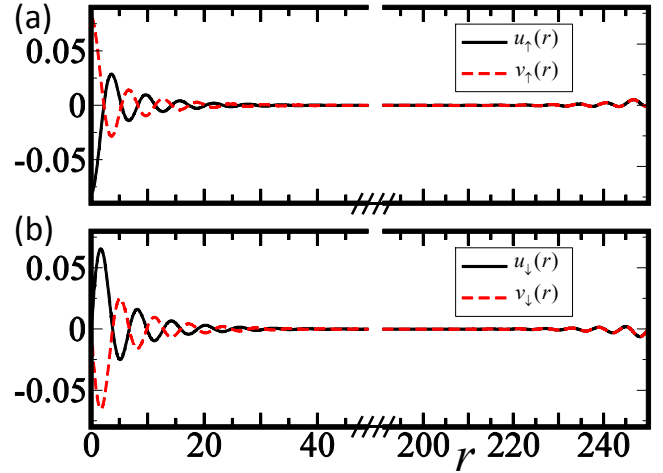} \vspace{-0pt}
\caption{(Color online) Plot of the wavefunctions $u_{\protect\sigma }\left(
r\right) $, $v_{\protect\sigma }\left( r\right) $ of the Majorana zero
energy state. $\protect\mu =0$, $V_{z}=1$, $\protect\alpha =1$. The
wavefunctions have two parts. Around $r=0$ is the zero energy state in the
vortex core. Around the edge $r=R$ is the zero energy chiral edge state. In
the vortex core, $u_{\protect\sigma }\left( r\right) =-$ $v_{\protect\sigma %
}\left( r\right) $. At the edge, $u_{\protect\sigma }\left( r\right) =v_{%
\protect\sigma }\left( r\right) $.}
\label{wavefunction}
\end{figure}

In Fig. \ref{band}, we plot the quasiparticle energy $E_{nl}$ at different
angular momentum $l$ channels. We see only at the $l=0$ channel, there is a
unique Majorana zero energy solution. At non-zero $l$ channels, there are
two discrete energy levels: one is the edge state, the other is the vortex
core state. This can be clearly seen from the corresponding
eigenwavefunctions for these two energy levels at the $l=-1$ channel, which
are plotted in Fig. \ref{edgecorefun}. The continuous spectrum in Fig. \ref%
{band} corresponds to the bulk excitations. Inside the vortex core, the
first excited state above the zero energy mode is at the $l=-1$ channel and
the corresponding energy difference is called the minigap. The minigap is
the energy gap that protect the Majorana zero energy state from finite
temperature and disorder effects, therefore its magnitude is crucially
important for the observation of non-Abelian statistics of quasiparticles in
this system.

The non-Abelian topological property of the Majorana zero energy states at
the $l=0$ channel originates from their special forms of the wavefunctions $%
u_{\sigma }(r)$, $v_{\sigma }(r)$, which lead to a self-Hermitian Bogoliubov
quasiparticle operator, $\gamma _{0}^{\dag }=\gamma _{0}$. Specifically, the
quasiparticle wavefunctions around the vortex core satisfy $u_{\sigma
}(r)=-v_{\sigma }(r)$, as clearly seen from Fig. \ref{wavefunction}. Chosen
an artificial overall phase $e^{i\pi /2}$ for the wavefunction (\ref{wave}),
it is easy to show the Bogoliubov quasiparticle operator $\gamma _{0}$
defined in Eq. (\ref{Bog}) satisfies $\gamma _{0}^{\dag }=\gamma _{0}$. Note
that $\gamma _{0}$ cannot be the electron creation or annihilation operator
because it does not obey the anticommutation relation for electrons. Instead
, the quasiparticle defined by $\gamma _{0}$ is called a Majorana fermion.
The exchange statistics of two Majorana quasiparticle excitations in two
vortex cores are not the simple Fermi statistics: they can be either Abelian
or non-Abelian in the degenerate subspace spanned by the Majorana operators
in a set of vortices \cite{NayakRMP}.
\begin{figure}[t]
\includegraphics[width=0.9\linewidth]{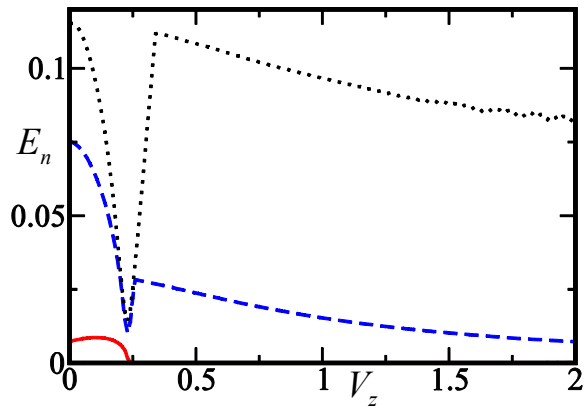} \vspace{0pt}
\caption{(Color online) Plot of three quasiparticle excitation energies $%
E_{n}$ with respect to the Zeeman field $V_{z}$ in a vortex core. Solid
line: ground state energy at $l=0$. Dashed line: ground state energy at $%
l=-1 $. Dotted line: the first excited state energy at $l=0$. $\protect%
\alpha =1$, $\protect\mu =0.2$. }
\label{EVZ}
\end{figure}

\begin{figure}[b]
\includegraphics[width=0.9\linewidth]{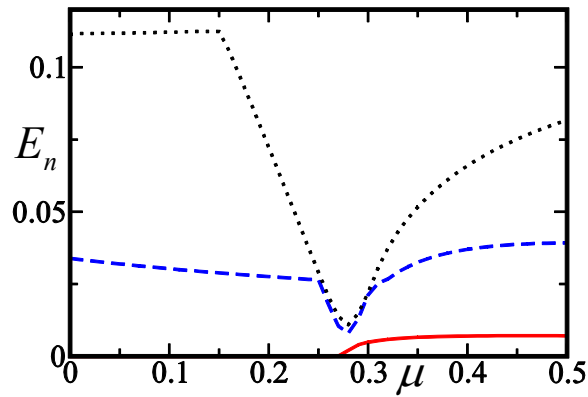} \vspace{-0pt}
\caption{(Color online) Plot of three quasiparticle excitation energies $%
E_{n}$ with respect to the chemical potential $\protect\mu $ in a vortex
core. The notation for different lines are the same as that in Fig. \protect
\ref{EVZ}. $\protect\alpha =1$, $V_{z}=0.3$.}
\label{ECHE}
\end{figure}

Interestingly, the wavefunctions still oscillate spatially even deep inside
the bulk region. Such an oscillation makes it difficult to realize a single
qubit gate for universal TQC using the tunneling between two vortices \cite%
{meng}. We also observe the zero energy edge states around the boundary,
which satisfy $u_{\sigma }(r)=v_{\sigma }(r)$, in contrast to $u_{\sigma
}(r)=-v_{\sigma }(r)$ in the vortex core. The wavefunctions $u_{\sigma }(r)$
and $v_{\sigma }(r)$ vanish at the boundary, as expected. An analytic
explanation for the observed $u_{\sigma }(r)=v_{\sigma }(r)$ relation of the
chiral edge modes is provided in Appendix C. We emphasize that zero energy
state in the vortex core and the edge must appear simultaneously because
Majorana zero energy modes only come in pairs (either between two vortices
or between a vortex and the edge).
\begin{figure}[t]
\includegraphics[width=0.9\linewidth]{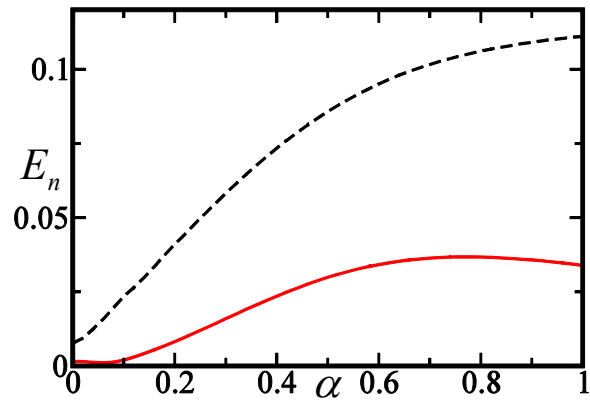} \vspace{-0pt}
\caption{(Color online) Plot of the bulk excitation gap (dashed line) and
the minigap (solid line) with respect to the spin-orbit coupling strength $%
\protect\alpha $. $\protect\mu =0$, $V_{z}=0.3$.}
\label{alpha}
\end{figure}

\begin{figure}[b]
\includegraphics[width=0.9\linewidth]{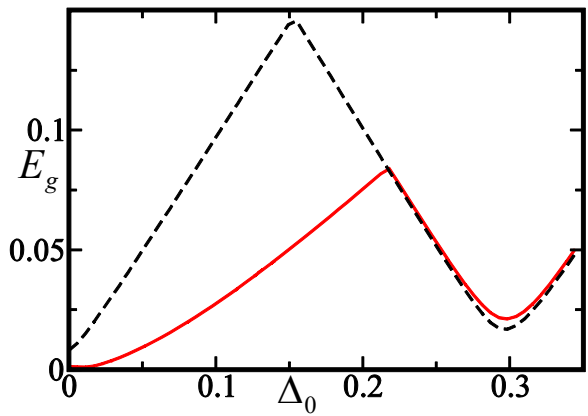} \vspace{-0pt}
\caption{(Color online) Plot of the minigap (solid line) and the bulk
excitation gap (dashed line) with respect to the uniform bulk
superconducting pairing gap $\Delta _{0}$. $\protect\mu =0$, $V_{z}=0.3$, $%
\protect\alpha =1$.}
\label{Egapdelta}
\end{figure}

In Figs. (\ref{EVZ},\ref{ECHE},\ref{alpha}), we plot three different
quasiparticle energies with respect to various physical parameters: (i) the
ground state energy in the vortex core at the $l=0$ channel. In certain
parameter region, this state is the zero energy Majorana mode; (ii) the
ground state energy in the vortex core at the $l=-1$ channel. In the
parameter region with the zero energy modes, this energy corresponds to the
minigap $E_{g}$; (iii) the first excited energy at the $l=0\,\ $channel. As
we can see from Fig. \ref{band}, it corresponds to the minimum bulk
excitation energy. Fig. \ref{EVZ} shows the dependence of these
quasiparticle energies on the Zeeman field $V_{z}$. We see the Majorana zero
energy state exists only in the region $V_{z}>\sqrt{\Delta _{0}^{2}+\mu ^{2}}%
\approx 0.23$. In this region, the minigap $E_{g}$ first increases rapidly
to a level $E_{g}^{\max }$, and then decreases slowly with increasing $V_{z}$%
. With a very large $V_{z}$, all electrons occupy the spin down states, and
there is no superconducting pairing. Therefore the decrease of the minigap
with increasing $V_{z}$ is expected. Note that $E_{g}$ is larger than the
typical minigap $E_{g}^{\prime }\sim \Delta _{0}^{2}\sim 0.01$ for a regular
(\textit{s}-wave or chiral \textit{p}-wave) superconductor/superfluid \cite%
{gap}. To characterize the dependence of the minigap $E_{g}$ on the uniform
bulk pairing gap $\Delta _{0}$, we numerically calculate $E_{g}$ for
different $\Delta _{0}$ and plot it in Fig. \ref{Egapdelta}. We see in the
region $\Delta _{0}=0.07\sim 0.22$, $E_{g}$ is roughly proportional to $%
\Delta _{0}$. When $\Delta _{0}>0.22$, the minigap decreases because $\Delta
_{0}$ is now very close to $V_{z}=0.3$. In general, $E_{g}$ may be a linear
combination of $\Delta _{0}$ and $\Delta _{0}^{2}$.

In Fig. \ref{ECHE}, we see the zero energy modes disappear in the region $%
\mu >\sqrt{V_{z}^{2}-\Delta _{0}^{2}}$. As expected, the minigap $E_{g}$ has
a maximum at $\mu =0$. $E_{g}$ only changes slightly when the chemical
potential varies. In Fig. \ref{alpha}, we plot the quasiparticle energy $%
E_{n}$ with respect to the spin-orbit coupling strength $\alpha $. We see
the minigap has a strong dependence on $\alpha $. $E_{g}$ initially
increases quickly with a growing $\alpha $, and reaches the maximum, then
decreases very slowly for large $\alpha $. This is expectable because the
Rashba spin-orbit coupling provides the necessary chirality for the zero
energy states. When $\alpha =0$, the coupling between spin up and down
states vanishes, and a pure $s$-wave superconductor does not have the zero
energy modes.

\begin{figure}[t]
\begin{center}
\includegraphics[width=0.9\linewidth]{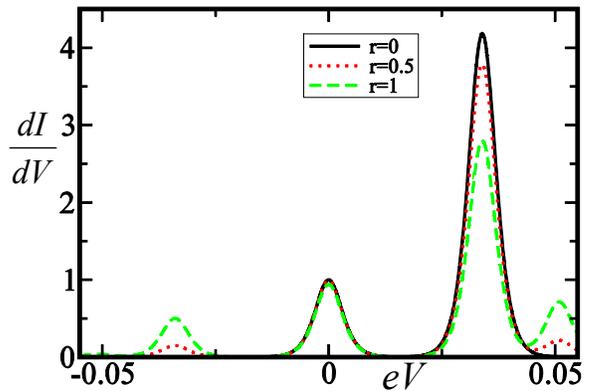}
\end{center}
\caption{(Color online) Plot of the STM tunneling conductance at different $%
r $ around the vortex. $\protect\alpha =1$, $\protect\mu =0$, $V_{z}=0.3$, $%
k_{B}T=\Delta _{0}/50$. }
\label{STM}
\end{figure}

The zero energy modes in the vortex core can be probed by bringing a STM tip
close to the vortex core in experiments. The resulting tunneling conductance
can be written as\cite{Gygi}
\begin{equation}
\frac{dI}{dV}\propto -\sum_{i\sigma }[u_{i\sigma }^{2}(r)f^{\prime
}(E_{i}-eV)+v_{i\sigma }^{2}(r)f^{\prime }(E_{i}+eV)],  \label{STMcond}
\end{equation}%
where $eV$ is the voltage bias, $i$ represents different energy level, $I$
is the tunneling current, $T$ is the operation temperature of the STM, $%
f=1/(\exp (\left( E_{i}-eV\right) /k_{B}T)+1)$ is the Fermi-Dirac
distribution, and the derivative of $f$ is with respect to $E$. In Fig. \ref%
{STM}, we plot the STM tunneling conductances at different radius of the
vortex core, which show clear zero energy peaks coming from the zero energy
states. The other peaks correspond to other vortex core states at the
angular momentum $l\neq 0$ channels. The first peaks around the zero energy
are at the $l=-1$ channels. The distance between the peak centers at $l=0$, $%
-1$ channels is a measurement of the minigap. There is also an asymmetry of
the peak strengths at $\pm eV$. Because of the symmetry $E\rightarrow -E$, $%
u_{l\sigma }(r)\leftrightarrow v_{-l\sigma }(r)$ in the BdG equation, we
only need the $E_{i}\geq 0$ terms in Eq. (\ref{STMcond}). Therefore the
magnitudes of the peaks at positive or negative $eV$ are proportional to $%
\sum_{\sigma }u_{i\sigma }^{2}(r)$ and $\sum_{\sigma }v_{i\sigma }^{2}(r)$
respectively. Generally, $\sum_{\sigma }u_{i\sigma }^{2}(r)\neq \sum_{\sigma
}v_{i\sigma }^{2}(r)$, leading to the asymmetric peaks around $eV=0$. In
particular, we find the peak for $r=0$ at the negative $eV$ disappears,
which means $\sum_{\sigma }v_{i\sigma }^{2}(0)$ must be zero for the minigap
state (\textit{i.e.}, $l=-1$, $E_{-1}>0$ state). As we can see from Eq. (\ref%
{BdG10}), $\sum_{\sigma }v_{i\sigma }^{2}(r)$ in the minigap state involves
the basis states $\phi _{l=-2,j}\left( r\right) $ and $\phi _{l=-1,j}\left(
r\right) $ which are zero at $r=0$ because the Bessel function $J_{l\neq
0}\left( r=0\right) =0$.

\begin{figure}[t]
\includegraphics[width=0.9\linewidth]{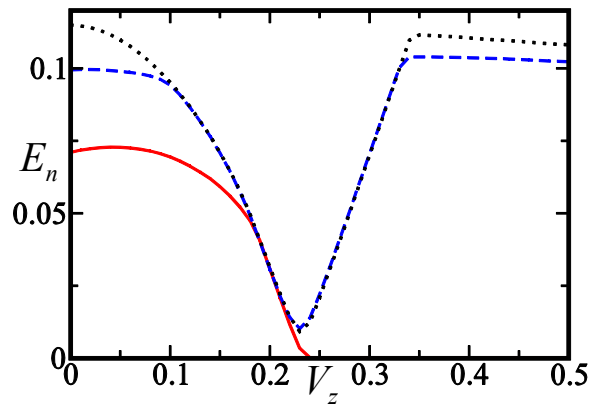} \vspace{0pt}
\caption{(Color online) Plot of three quasiparticle excitation energies $%
E_{n}$ with respect to the Zeeman field $V_{z}$ in the vortex core in the
presence of a Gaussian impurity. The notation for different lines are the
same as that in Fig. \protect\ref{EVZ}. $\protect\mu =0.2$, $\protect\alpha %
=1$. $U_{0}=10$, $s=0.5$.}
\label{EimpuG}
\end{figure}

\section{Effects of impurities}

In a realistic TQC platform, both Majorana zero energy states and minigaps
need be robust in the presence of impurities. A general argument on the
robustness of zero energy states is based on the particle-hole symmetry in
the BdG equation, which ensures its energy spectrum to be symmetric upon $%
E\rightarrow -E$, that is, the $E$ and $-E$ states must come in pairs.
However, the zero energy does not come in pairs in a BdG equation (see Fig. %
\ref{band}). Therefore a local small perturbation cannot destroy this
symmetry and shifts the zero energy to a finite energy $E$ that must emerge
simultaneously with another state with an energy $-E$. Currently, the
robustness of the magnitude of the minigap in the presence of impurities is
still not clear. In the following, we consider three types of rotationally
invariant impurities and study their effects on the zero energy states and
the minigaps in the vortex core. Interestingly, we find that impurities may
actually increase the minigaps in the vortex.

\begin{figure}[t]
\includegraphics[width=0.9\linewidth]{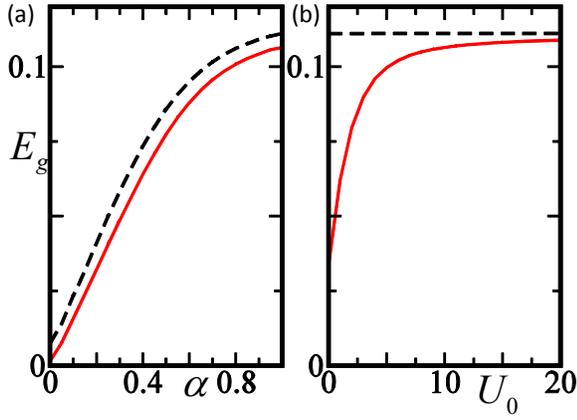} \vspace{0pt}
\caption{(Color online) Plot of the bulk excitation gap (dashed line) and
the minigap (solid line) with respect to the spin-orbit coupling strength $%
\protect\alpha $ (a) and the impurity strength $U_{0}$ (b) in the presence
of a Gaussian impurity potential. $V_{z}=0.3$, $\protect\mu =0$, $s=0.5$.
(a) $U_{0}=10$. (b) $\protect\alpha =1$.}
\label{Eimpu-gap}
\end{figure}

\begin{figure}[b]
\includegraphics[width=1.0\linewidth]{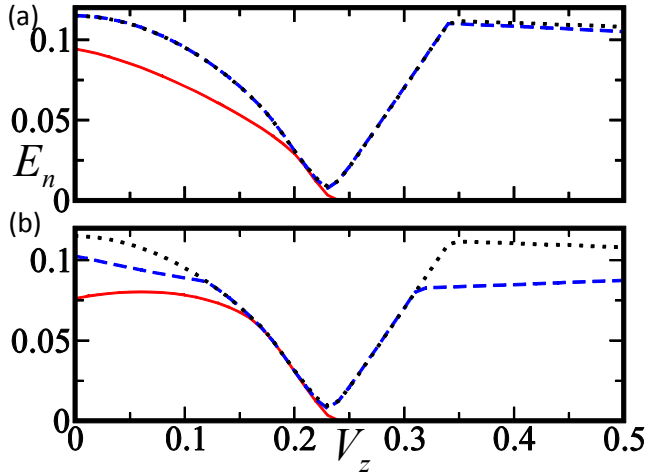} \vspace{0pt}
\caption{(Color online) Plot of three quasiparticle excitation energies $%
E_{n}$ with respect to the Zeeman field $V_{z}$ in the vortex core in the
presence of a magnetic Gaussian impurity. The notation for different lines
are the same as that in Fig. \protect\ref{EVZ}. $\protect\mu =0.2$, $\protect%
\alpha =1$, $s=0.5$. (a) $U_{0}=10$; (b) $U_{0}=-10$. }
\label{EimpuGz}
\end{figure}

First, we consider a spin-independent Gaussian impurity
\begin{equation}
U(r)=U_{0}\exp \left( -r^{2}/2s^{2}\right)  \label{impurity1}
\end{equation}%
in the BdG equation (\ref{BdG4}) for a vortex, where $U_{0}$ is the impurity
strength, $s$ is the half-width of the impurity potential that is comparable
to the size of the vortex core. In Fig. \ref{EimpuG}, we plot the
quasiparticle energies with respect to $V_{z}$ in the presence of a Gaussian
impurity (\ref{impurity1}). In a regular \textit{s}-wave superconductor, it
is expected that such an impurity potential can couple different
quasiparticle excitation states, and thus modify the energy spectrum. This
can be seen in Fig. \ref{EimpuG} in the parameter region $V_{z}<\sqrt{\mu
^{2}+\Delta _{0}^{2}}$ without the zero energy states. We find that the zero
energy states in the region $V_{z}>\sqrt{\mu ^{2}+\Delta _{0}^{2}}$ are very
robust even for a large impurity potential $U_{0}=10$. More interestingly,
we find that, by comparing with Fig. \ref{EVZ}, the presence of the Gaussian
impurity potential can increase the magnitude of the minigaps significantly.
In Fig. \ref{Eimpu-gap}, we plot the minigap and the bulk excitation gap in
the presence of a Gaussian impurity potential. We see the minigap approaches
the bulk excitation gap for a large $U_{0}$. This enhancement of the minigap
can be understood by considering the fact that the impurity potential can
repulse (or attract) electrons and enlarge the energy level splitting
between different discrete states in the vortex core. In practice, we may
add a Gaussian type of potential at the vortex center to increase the
minigaps and ensure the corresponding topological protection from finite
temperature effects.

\begin{figure}[t]
\includegraphics[width=0.9\linewidth]{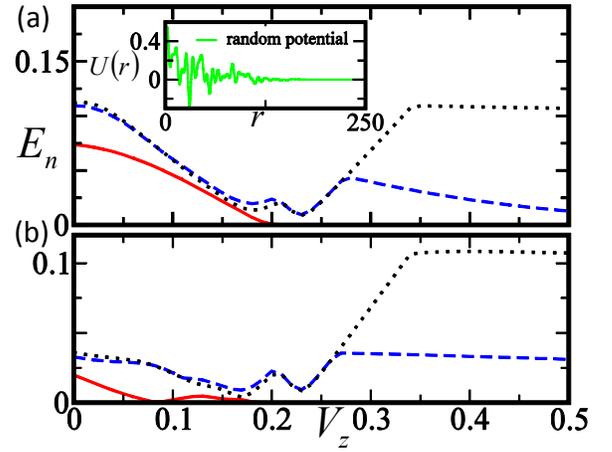} \vspace{0pt}
\caption{(Color online) Plot of three quasiparticle excitation energies $%
E_{n}$ with respect to the Zeeman field $V_{z}$ in the vortex core in the
presence of a Gaussian random impurity (a) and a magnetic Gaussian random
impurity (b). The notation for different lines are the same as that in Fig.
\protect\ref{EVZ}. $\protect\mu =0.2$, $\protect\alpha =1$. $U_{0}=1$, $%
s=0.5 $. }
\label{Eimpuran}
\end{figure}

Secondly, we consider a magnetic impurity potential $U(r)\sigma _{z}$, whose
effects on the zero energy modes and minigaps are shown in Fig. \ref{EimpuGz}%
. Such an impurity potential acts as a spatially localized Zeeman field. No
matter the local Zeeman field $U(r)\sigma _{z}$ is along the same (Fig. \ref%
{EimpuGz}a) or opposite (Fig. \ref{EimpuGz}b) direction as the original
Zeeman field $V_{z}$, the zero energy states do not change and the minigaps
are enhanced significantly. Note that in both cases the impurity does have a
significant impact on the quasiparticle energies in the region $V_{z}<\sqrt{%
\mu ^{2}+\Delta _{0}^{2}}$.

Finally, we consider random impurity potentials
\begin{equation}
U_{ran}(r)=\sum_{m=1}^{N}U_{0}\sin \omega _{m}r\exp \left(
-r^{2}/2s^{2}\right)  \label{impurity2}
\end{equation}%
and $U_{ran}(r)\sigma _{z}$, where $\omega _{m}$ are random frequencies (see
the inset in Fig. \ref{Eimpuran}a for the spatial profile of the random
potential). This potential still has significant amplitudes outside the
vortex core region. In Fig. \ref{Eimpuran}, we find that the critical $V_{z}$
for the existence of the zero energy modes is smaller than the expected $%
\sqrt{\mu ^{2}+\Delta _{0}^{2}}$. It may originate from the long range
property of the random potential, whose average $\bar{U}$ around the vortex
core shifts the chemical potential $\mu $. Therefore the critical point is
shifted to $\sqrt{(\mu -\bar{U})^{2}+\Delta _{0}^{2}}$. For a magnetic
random impurity $U_{ran}(r)\sigma _{z}$, the corresponding critical point
shifts to $\sqrt{\mu ^{2}+\Delta _{0}^{2}}-\bar{U}$. However, the minigaps
are small in these parameter regions, therefore they cannot be used as a TQC
platform even the zero energy states exist. Only in the region $V_{z}>\sqrt{%
\mu ^{2}+\Delta _{0}^{2}}$, the zero energy states and the minigaps are
robust against impurities.

So far the impurity potentials have been chosen to be rotationally invariant
to simplify the numerical calculation. Non-rotationally invariant potentials
couple different angular momentum channels and thus are difficult to
simulate numerically. In a realistic situation, vortices in superconductors
are often pinned by impurities so that the centers of the impurities and
vortices are the same to minimize the free energy of the system. Such
overlapping of the centers provides one justification of our choice of
rotationally invariant impurity potentials. More generally, an impurity
potential may be expanded as
\begin{equation}
U(r,\theta )=\sum_{n=-\infty }^{\infty }U_{n}(r)\exp (in\theta ),
\label{impgen}
\end{equation}%
where $U_{0}(r)$ is the rotationally invariant part, while the $U_{n\neq
0}(r)$ corresponds to the non-rotationally invariant part. For instance, a
Gaussian impurity potential located at $(x_{0},0)$ instead of $(0,0)$ can be
written as $U(\mathbf{r})=\exp \left( -\frac{r^{2}+x_{0}^{2}}{2s^{2}}\right)
\exp \left( \frac{rx_{0}}{s^{2}}\cos \theta \right) =\exp \left( -\frac{%
r^{2}+x_{0}^{2}}{2s^{2}}\right) \sum_{n=-\infty }^{\infty }I_{n}(\frac{rx_{0}%
}{s^{2}})\exp (in\theta )$, where $I_{n}$ is the modified Bessel functions.
Because the eigenstates of the BdG equation (\ref{BdG4}) have fixed angular
momentum $l$, $U_{0}(r)$ only couples states with the same $l$, while $%
U_{n\neq 0}(r)$ couples states with different $l$. We can treat the impurity
potential as a perturbation to the original BdG Hamiltonian. The first order
energy correction to the zero energy state is $\Delta
E_{0}^{(1)}=\sum_{\sigma }\left\langle u_{0\sigma }\right\vert U(r,\theta
)\left\vert u_{0\sigma }\right\rangle -\left\langle v_{0\sigma }\right\vert
U(r,\theta )\left\vert v_{0\sigma }\right\rangle =0$ because $u_{0\sigma
}=-v_{0\sigma }$ around the vortex. The second order correction $\Delta
E_{0}^{(2)}=-\sum_{i}\left\vert U_{0i}\right\vert ^{2}/E_{i}=0$ because of
the coexistence of $E_{i}$ and $-E_{i}$ in the energy spectrum. Therefore
the zero energy states are preserved by the particle hole symmetry in the
BdG equation, as discussed at the beginning of this section. As to the
minigap state, we see the first order correction $\Delta E_{1}^{(1)}=$ $%
\left\langle \Phi _{1}^{l=-1}(\mathbf{r})\right\vert U_{0}(r)\left\vert \Phi
_{1}^{l=-1}(\mathbf{r})\right\rangle $ only depends on the rotationally
invariant part. The non-rotationally invariant part $U_{n\neq 0}(r)$ only
yields a second or higher order correction to the minigap, which is
generally small and may be neglected.

\section{Conclusion}

In conclusion, we show that Majorana zero energy states and minigaps of
quasiparticle excitations in a vortex are robust in a heterostructure
composed of an \textit{s}-wave superconductor, a spin-orbit-coupled
semiconductor thin film, and a magnetic insulator. By numerically solving
BdG equations with a vortex in a cylinder geometry, the dependence of
Majorana zero energy states and minigaps on various physics parameters of
the system (Zeeman field, chemical potential, spin-orbit coupling strength)
is characterized. In certain parameter region, the minigap is proportional
to the pairing gap $\Delta _{s}$, instead of $\sim \Delta _{s}^{2}$ for a
regular chiral $p$-wave superconductor/superfluid. The existence of the zero
energy chiral edge state at the boundary is demonstrated both analytically
and numerically. The STM tunneling conductance in the vortex core shows
pronounced peaks at the zero energy as well as the minigap energy, which can
be used to measure the zero energy state and the minigap. We show that the
Majorana zero energy states are robust in the presence of various types of
impurities. Surprisingly, we find that impurity potentials may greatly
enhance the magnitudes of the minigaps. Therefore they can be induced
intentionally in the heterostructure to enhance the topological protection
of the non-Abelian platform. We believe our characterization of the
behaviors of the zero energy modes and the minigaps in different parameter
regions will help the design of a realistic semiconductor-superconductor
heterostructure system for the experimental observation of non-Abelian
statistics and the physical implementation of TQC in the future.

\textbf{Acknowledgement:} We thank Sumanta Tewari for helpful discussion.
This work is supported by the DARPA-YFA (N66001-10-1-4025), DARPA-MTO (FA9550-10-1-0497), and the ARO
(W911NF-09-1-0248).

\appendix

\section{Numerical method for solving the BdG equation}

We expand the radial wavefunctions $u_{n\sigma }^{l}(r),v_{n\sigma }^{l}(r)$
in Eq. (\ref{wave}) using the basis states $\phi _{lj}(r)=\sqrt{2}%
J_{l}(\beta _{lj}r/R)/[RJ_{l+1}(\beta _{lj})]$, where $\beta _{lj}$ is the $%
j $-th zero of the Bessel function $J_{l}(x)$. The BdG eigenvalue equation (%
\ref{BdG4}) reduces to a block diagonal matrix
\begin{equation}
\left(
\begin{array}{cccc}
T_{l}^{+} & S_{l} & \Delta _{l} & 0 \\
S_{l}^{T} & T_{l+1}^{-} & 0 & \Delta _{l+1} \\
\Delta _{l}^{T} & 0 & -T_{l-1}^{-} & -S_{l-1} \\
0 & \Delta _{l+1}^{T} & -S_{l-1}^{T} & -T_{l}^{+}%
\end{array}%
\right) \left(
\begin{array}{c}
u_{n\uparrow }^{l} \\
u_{n\downarrow }^{l+1} \\
v_{n\downarrow }^{l-1} \\
-v_{n\uparrow }^{l}%
\end{array}%
\right) =E_{nl}\left(
\begin{array}{c}
u_{n\uparrow }^{l} \\
u_{n\downarrow }^{l+1} \\
v_{n\downarrow }^{l-1} \\
-v_{n\uparrow }^{l}%
\end{array}%
\right) ,  \label{BdG10}
\end{equation}%
where
\begin{eqnarray}
(T_{l}^{\pm })_{ij} &=&(\frac{\beta _{lj}^{2}}{R^{2}}\pm V_{z}-\mu )\delta
_{ij},  \label{T} \\
(S_{l})_{ij} &=&\alpha \int_{0}^{R}r\phi _{li}(r)(\partial _{r}+\frac{l+1}{r}%
)\phi _{l+1j}(r)dr, \\
(\Delta _{l})_{ij} &=&\int_{0}^{R}r\Delta _{s}\left( r\right) \phi
_{li}(r)\phi _{l-1j}(r)dr, \\
u_{n\sigma }^{l} &=&[u_{n\sigma 1}^{l},\cdot \cdot \cdot ,u_{n\sigma
j}^{l},\cdot \cdot \cdot ]^{T}, \\
v_{n\sigma }^{l} &=&[v_{n\sigma 1}^{l},\cdot \cdot \cdot ,v_{n\sigma
j}^{l},\cdot \cdot \cdot ]^{T}.
\end{eqnarray}

\section{Finite size effect}

In an infinite large system, the zero energy modes always exist in the
parameter region $V_{z}>\sqrt{\Delta ^{2}+\mu ^{2}}$ \cite{Sau10}. In Ref.
\cite{Sau10}, the existence of the zero energy modes was proven by solving
the BdG equation separately in and out the vortex core edge and matching the
boundary condition at the vortex edge. A step function for the pairing gap
has been used in this case to simplify the calculation. The zero energy
modes exist when the number of unknown coefficients of the wavefunctions is
equal to the number of independent constraints (vortex edge boundary
conditions and normalization of the wavefunction), which occurs in the
parameter region $V_{z}>\sqrt{\Delta ^{2}+\mu ^{2}}$. However, there will be
additional constraints in the system boundary, therefore zero energy
solutions do not exist generally in a finite size system, that is, the
quantum well state may destroy the zero energy modes when the size of the
material or the confining potential is small enough.

\begin{figure}[t]
\includegraphics[width=0.9\linewidth]{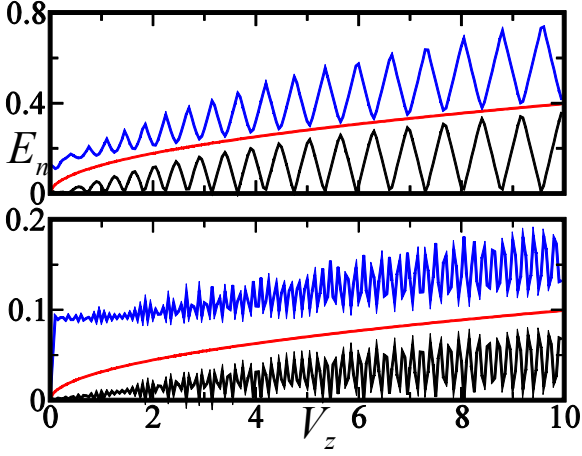} \vspace{0pt}
\caption{(Color online) Plot of the first two quasiparticle energies at the $%
l=0$ channel with respect to the Zeman field $V_{z}$. $\protect\mu =0$, $%
\protect\alpha =1$. (a) $R=25$; (b) $R=100$.}
\label{qse}
\end{figure}

To describe the finite size effect, we consider the parameter region $%
V_{z}\rightarrow \infty $, where the spin-orbit coupling and proximity
induced superconductivity are not important. In this region, the BdG matrix (%
\ref{BdG10}) is diagonal, and the zero energy solution exists when
\begin{equation}
(T_{1}^{-})_{nn}=\frac{\beta _{1n}^{2}}{R^{2}}-V_{z}=0.
\end{equation}%
Here we take $\mu =0$. We see there are zero energy modes only at some
special $V_{z}=\beta _{1n}^{2}/R^{2}$. When $V_{z}$ exceeds $\beta
_{1n}^{2}/R^{2}$, the lowest two eigenvalues are $V_{z}-\beta
_{1n}^{2}/R^{2} $ and $\beta _{1n+1}^{2}/R^{2}-V_{z}$, and the lowest energy
branch reaches its local maximum at
\begin{equation}
\frac{\beta _{1n+1}^{2}}{R^{2}}-V_{z}=V_{z}-\frac{\beta _{1n}^{2}}{R^{2}},
\end{equation}%
where the integer $n$ is determined by
\begin{equation}
\frac{\beta _{1n}^{2}}{R^{2}}<V_{z}<\frac{\beta _{1n+1}^{2}}{R^{2}}.
\end{equation}%
$\beta _{1n}$ is the $n$-th zero of the Bessel function $J_{1}\left(
x\right) $ and satisfies $\beta _{1n+1}\approx \beta _{1n}+\pi $, therefore
the maximum energy is
\begin{equation}
E_{max}=\frac{\beta _{1n+1}^{2}-\beta _{1n}^{2}}{2R^{2}}\approx \frac{\pi
\sqrt{V_{z}}}{R}.
\end{equation}

As an example, we consider the parameter $\alpha =1$, $\mu =0$, $l=0$ and
use the pairing gap from the \textit{s}-wave superconductor. In Fig. (\ref%
{qse}), we plot the lowest two quasiparticle energies with respect to $V_{z}$
for two different sizes of the system $R=25,100$. We see when $V_{z}$ is big
enough, the fitting function ($E_{max}=\pi \sqrt{V_{z}}/R$) is a good
approximation to the oscillation amplitude of the ground state. While at a
small $V_{z}$, the amplitude is much smaller than $\pi \sqrt{V_{z}}/R$. It
is clear when $R$ is big, the oscillation amplitude is strongly suppressed.
In practice, we choose the parameters $R=250$, $V_{z}<2$, and the finite
size effects can barely be seen and thus be neglected, as clearly
demonstrated in Fig. (\ref{EVZ}).

\section{Chiral edge states at the $l=0$ channel}

At the $l=0$ channel, the BdG equation for the zero energy state can be
written as

\begin{equation}
\left(
\begin{array}{cc}
H_{0} & \Delta _{s}(r) \\
\Delta _{s}^{\ast }(r) & -\sigma _{y}H_{0}^{\ast }\sigma _{y}%
\end{array}%
\right) \Phi _{0}(r)=0,  \label{BdG5}
\end{equation}%
where%
\begin{equation}
H_{0}=\left(
\begin{array}{cc}
\digamma \left( r\right) +V_{z}-\mu & \alpha \left( \partial _{r}+\frac{1}{r}%
\right) \\
-\alpha \partial _{r} & \digamma \left( r\right) +\frac{\eta }{r^{2}}%
-V_{z}-\mu%
\end{array}%
\right) ,  \label{Ham2}
\end{equation}%
$\digamma \left( r\right) =-\eta \left( \partial _{r}^{2}+\frac{1}{r}%
\partial _{r}\right) $. The BdG equation can be further reduced to a $%
2\times 2$ matrix differential equation

\begin{equation}
\left(
\begin{array}{cc}
\digamma \left( r\right) +V_{z}-\mu & \lambda \Delta _{s}\left( r\right)
+\alpha \left( \partial _{r}+\frac{1}{r}\right) \\
-\lambda \Delta _{s}\left( r\right) -\alpha \partial _{r} & \digamma \left(
r\right) +\frac{\eta }{r^{2}}-V_{z}-\mu%
\end{array}%
\right) \Psi _{0}(r)=0,  \label{BdG6}
\end{equation}%
where the parameter $\lambda =\pm 1$ and $i\sigma _{y}\tau _{y}\Phi
_{0}(r)=i\lambda \Phi _{0}(r)$ for nondegenerate zero energy states. These
conditions yield $\Psi _{0}(r)=\left[ u_{\uparrow },u_{\downarrow }\right]
^{T}$ and $u_{\sigma }(r)=\lambda v_{\sigma }(r)$. It has been shown in Ref.
\cite{Sau10} that only in the parameter region $\lambda =-1$, $V_{z}>\sqrt{%
\Delta _{0}^{2}+\mu ^{2}}$, there is an unique zero energy solution. Here we
show that a zero energy chiral edge state exists in the parameter region $%
\lambda =1$, $V_{z}>\sqrt{\Delta _{0}^{2}+\mu ^{2}}$.

As we can see from Fig. \ref{delta}, the pairing gap $\Delta _{s}(r)$
decreases from its bulk value $\Delta _{0}$ to zero at the boundary. For
simplicity, we approximate the radial dependence $\Delta _{s}(r)$ around the
boundary with a step function: $\Delta \left( r\right) =0$ for $R>r>R-\delta
$, and $\Delta \left( r\right) =\Delta _{0}$ for $r\leq R-\delta $, where $%
\delta $ is the coherence length of $\Delta _{s}(r)$ around the edge.
Because $R$ is very large, the BdG equation (\ref{BdG6}) reduces to
\begin{equation}
\left(
\begin{array}{cc}
-\eta \partial _{r}^{2}+V_{z}-\mu & \lambda \Delta _{s}\left( r\right)
+\alpha \partial _{r} \\
-\lambda \Delta _{s}\left( r\right) -\alpha \partial _{r} & -\eta \partial
_{r}^{2}-V_{z}-\mu%
\end{array}%
\right) \Psi _{0}(r)=0  \label{BdG7}
\end{equation}%
around the boundary. In\ the region $R>r>R-\delta $, the solution of Eq. (%
\ref{BdG7}) is given by $\Psi _{0}(r)=\left[ u_{\uparrow },u_{\downarrow }%
\right] ^{T}\exp \left( z\left( r-R\right) \right) $ with the constraint
\begin{equation}
\left(
\begin{array}{cc}
-\eta z^{2}+V_{z}-\mu & z\alpha \\
z\alpha & -\eta z^{2}-V_{z}-\mu%
\end{array}%
\right) \left(
\begin{array}{c}
u_{\uparrow } \\
u_{\downarrow }%
\end{array}%
\right) =0.  \label{BdG8}
\end{equation}%
The characteristic equation for $z$ is $\left( \eta z^{2}+\mu \right)
^{2}-V_{z}^{2}-z^{2}\alpha ^{2}=0$. There are four solutions of Eq. (\ref%
{BdG7}) which are well behaved at the edge: $\phi _{i}(r)=\left[ u_{\uparrow
}^{i},u_{\downarrow }^{i}\right] ^{T}exp\left( z_{i}\left( r-R\right)
\right) $ ($i=1,2,3,4$), where $z_{1}$, $z_{2}$, $z_{3}$, $z_{4}$, are four
solutions of Eq. (\ref{BdG8}). The full wavefunction in the region $%
R>r>R-\delta $ can be written as $\Psi _{0}(r)=c_{1}\phi _{1}(r)+c_{2}\phi
_{2}(r)+c_{3}\phi _{3}(r)+c_{4}\phi _{4}(r)$.

Far from the boundary, where $\Delta \left( r\right) =\Delta _{0}$, we can
expand the solution as a series in $\frac{1}{R-r}$%
\begin{equation}
\Psi _{0}(r)=\frac{\exp \left( iz\left( R-r\right) \right) }{\sqrt{R-r}}%
\sum_{n=0,1,2,...}\frac{a_{n}}{\left( R-r\right) ^{n}}  \label{Ser}
\end{equation}%
where $a_{n\text{ }}$are the corresponding spinors. The zeroth order
coefficient $a_{0}$ satisfies the following equation:%
\begin{equation}
\left(
\begin{array}{cc}
\eta z^{2}+V_{z}-\mu & \lambda \Delta _{0}-iz\alpha \\
-\lambda \Delta _{0}+iz\alpha & \eta z^{2}-V_{z}-\mu%
\end{array}%
\right) a_{0}=0.  \label{BdG9}
\end{equation}%
The higher order coefficients $a_{n\text{ }}$can be calculated from $a_{0}$
using a set of recursion relations. The characteristic equation has 4
complex root for $z$, same as that in Ref. \cite{Sau10}. Because Im$\left[
z_{n}\right] <0$ is required for a physical solution, there are three
independent roots only in the parameter region $\lambda =1$, $V_{z}>\sqrt{%
\Delta _{0}^{2}+\mu ^{2}}$, which yields three independent coefficients.
Together with the four independent coefficients $c_{1}$, $c_{2}$, $c_{3}$, $%
c_{4}$, in the region $R>r>R-\delta $, and seven constraints (match of $\
\Psi _{0}(r)$ and $\Psi _{0}^{\prime }(r)$ at $r=R-\delta $, the boundary
condition $\Psi _{0}(R)=0$, and the normalization of the wavefunction), we
can obtain a unique zero energy edge state at the boundary. However, the
chiral edge state wavefunction satisfies $u_{\sigma }(r)=v_{\sigma }(r)$
instead of $u_{\sigma }(r)=-v_{\sigma }(r)$ for the vortex core state, as
clearly seen from Fig. \ref{wavefunction}.

\end{document}